# Modelling brain tissue elasticity with the Ogden model and an alternative family of constitutive models†

**Afshin Anssari-Benam[1,*], Michel Destrade[2,3] and Giuseppe Saccomandi[2,4]**

[1]*Cardiovascular Engineering Research Lab (CERL), School of Mechanical and Design Engineering, University of Portsmouth, Anglesea Road, Portsmouth PO1 3DJ, United Kingdom. ORCID: 0000-0003-4348-5186.*

[2]*School of Mathematical and Statistical Sciences, NUI Galway, University Road, Galway, Ireland.*

[3]*Key Laboratory of Soft Machines and Smart Devices of Zhejiang Province and Department of Engineering Mechanics, Zhejiang University, Hangzhou 310027, PR China. ORCID: 0000-0002-6266-1221*

[4]*Dipartimento di Ingegneria, Università degli studi di Perugia, Via G. Duranti, Perugia 06125, Italy. ORCID: 0000-0001-7987-8892.*

## Summary

The Ogden model is often considered as a standard model in the literature for application to the deformation of brain tissue. Here we show that, in some of those applications, the use of the Ogden model leads to non-convexity of the strain-energy function and mis-prediction of the correct concavity of the experimental stress-stretch curves over a range of the deformation domain. By contrast, we propose a family of models which provides a favourable fit to the considered datasets while remaining free from the highlighted shortcomings of the Ogden model. While, as we discuss, those shortcomings might be due to the artefacts of the testing protocols, the proposed family of models proves impervious to such artefacts.

## 1. Introduction

Studying the mechanical behaviour of brain tissue evinces interesting challenges from the perspective of experimental and constitutive modelling [1,2]. At the macro level, brain tissue is extremely soft and highly heterogeneous [1,2]. At the microstructural level, the tissue consists of two distinct grey and white matter regions with different microstructures [3,4]. This heterogeneity, coupled with the extreme softness of the tissue, provides a formidable challenge to preparing test specimens that are truly a representative volume element of the whole tissue, as well as minimising artefacts such as gripping effects, friction, etc. Some studies attribute the wide variation in the existing experimental data found in the literature to the structural heterogeneity of the specimens in each study; see, e.g., [5,6].

Despite these challenges, many studies have attempted to characterise the mechanical behaviour of brain tissue with the aim of establishing constitutive relationships between mechanical stimuli and the ensuing tissue-level responses.

† Celebrating 50 years of Professor Ray Ogden's seminal model in nonlinear elasticity.

*Author for correspondence (email: afshin.anssari-benam@port.ac.uk).

Tissue mechanics has been shown to play a key role in regulating the development of the brain and maintaining its homeostasis, as well as influencing its pathology; e.g., an increase in the brain tissue stiffness has been shown to enhance the migration and proliferation of tumour cells [7], see also Barnes et al. [8] and Goriely et al. [2] for a comprehensive review. These findings underline the significance of properly characterising the mechanical properties and behaviour of brain tissue. However, the complexity of the mechanics of brain tissue in normal function and pathology extends beyond what may be achieved in a laboratory setting *in vitro*. Mathematical and computational models therefore become essential investigatory tools in gaining insight into the complex mechanical behaviour, properties and function of the brain.

The suitability of constitutive models is judged based on how accurately they capture and predict the various aspects of the mechanical properties and behaviours of the tissue, as revealed by experiments *in vitro*. These mathematical models then form a core for computational modelling and *in silico* simulations of the physiology, function and pathology of the brain under more realistic and complex physiological and pathological conditions. The fidelity of the computational analyses is therefore highly reliant on the suitability of the choice of the mathematical model, which in turn is informed by the mechanical behaviour observed in the experiments [5,6].

Of the various mechanical loading experiments that may be employed to inform and calibrate an appropriate choice of an elastic model for the mechanics of brain tissue, particular emphasis has been afforded to the shear and compression deformations. Compression is the most relevant deformation mode for common traumatic brain injuries, and shear strains are thought to be the main mechanical cause of diffuse axonal injuries [5]. However, tensile deformation tests have also been carried out on brain tissue specimens, e.g., studies by Budday et al. [3,4]. Goriely et al. [2] identify the study by Estes and McElhaney [10] as the first to perform large deformation tests, up to 270% strain, on human and Rhesus monkey brain tissues, which revealed the nonlinear deformation of the tissue "with concave upward stress-strain curves". Recent studies have underlined more complex aspects of the mechanical behaviour of brain tissue including its time (rate) dependent [5,6] and biphasic nature (e.g., see [11] and also Budday et al. [1] and Goriely et al. [2] for reviews). A concise but informative summary of the timescale of the mechanical processes affecting brain tissue mechanics, in relation to choosing an appropriate modelling criterion, may be found in [12]. The general consensus, however, is that while the mechanics of brain tissue deformation in processes such as surgery may best be modelled through viscoelasticity and/or poroelasticity, and that fast processes are rate-dependent, we may nonetheless resort to hyperelasticity as the guiding approximation [1,3,4,12] for fixed intermediary loading rates (either fast or slow). In this manuscript, therefore, we only concern ourselves with the hyperelastic response, properties and modelling of brain tissue. Of course, we keep in mind the problem of reductionism in biomechanics, remaining aware that modelling of biological organs requires much more than hyperelasticity, since a realistic model of a tissue's mechanical response should encompass interactions of the extracellular matrix with physiological fluids, biochemical reactions characterising the physiological and/or pathological functions of the organ itself, feedback loops, etc.; see Rajagopal and Rajagopal [13] for a discussion on these points.

The key features of the hyperelastic behaviour of brain tissue as observed in the experimental results reported by Budday and collaborators [3,4,14] and Mihai et al. [15], may be summarised as follows:

a. The shear response is highly nonlinear, as is the compression-tension behaviour [3,4,15];

b. The compression-tension deformation is asymmetric [3,4];

c. It may be appropriate to consider brain tissue as isotropic and incompressible [14];

We will henceforth refer the above list as 'Features (a) to (c)'.

To capture these mechanical features, some studies have advocated the use of the one-term Ogden model; e.g., see [3,4], while others have also considered higher terms (from three- to eight-term Ogden models); see, e.g., [15]. What is intriguing here is that, as we shall see, the considered applications of the Ogden model as used in those studies all lead to problematic results such as the non-convexity of the iso-energy plots of the characterised strain energy function and/or an erroneous concavity of the predicted stress-train curves, at least over a range of the deformation domain. The former may lead to physical, material and numerical instabilities (see [16, 17] for extensive discussions) and the latter is not supported by the data.

The aim of this paper is to make the case for the use of a different family of models, based on a recently proposed parent model [18], to the mechanics of brain tissue by way of: (i) highlighting the aforementioned problems upon using the Ogden model; (ii) illustrating the capability of our proposed models to capture Features (a) to (c); and (iii) demonstrating the favourable fits to the experimental data obtained by the proposed models.

In §2 the mathematical formulation of the proposed models and a brief background to the development of the models is presented. This is followed in §3 by the demonstration of the suitability and capability of the proposed models in capturing Features (a) to (c) of the brain tissue mechanics via the application of the models to extant experimental compression-tension and simple shear datasets via simultaneous fitting. We find favourable agreements between the modelling predictions and the data. We also present a direct comparison with the results of the Ogden model, demonstrating that, as used in the specified studies, it leads to the issues of non-convexity of the strain-energy and/or non-concavity of the tensile response. In §4 we present a phenomenological generalisation of the proposed parent model [18] into a functional representation which embodies the principal stretches $\lambda_i$ instead of the principal invariants $I_i$. On using the datasets of Budday et al. [3,4], we show that this model fits the brain deformation data favourably, providing fits as good as the Ogden model while remaining convex. We provide concluding remarks in §5, with some further thoughts on the mechanics of brain tissue and possible experimental artefacts. We hypothesise that the problems exhibited by the Ogden model in the analysis of the brain tissue biomechanics may be due to the specifics of the simple shear experimental protocols chosen in studies and as such, may be resolvable. However, we stipulate, given that the proposed models herein prove to remain impervious to the ill-posed effects of the employed experimental protocols, their application to the biomechanics of brain is advantageous.

# 2. Modelling framework

In this section we provide a brief background on the underlying mechanical and mathematical motivations of the proposed family of models, and present the invariants-based models (namely, *Models I* and *II*). A third form of the model in this family, *Model III*, is presented in §4.

The proposed modelling framework which we wish to explore and utilise in this work centres on the *parent* strain energy function of the form:

$$W = \mu N \left[ \frac{1}{6N} (I_1 - 3) - \ln \left( \frac{I_1 - 3N}{3 - 3N} \right) \right], \tag{2.1}$$

where $\mu$ and $N$ are model parameters and $I_1$ is the first invariant of the left Cauchy-Green deformation tensor **B**, with the constraint $I_1 < 3N$ imposed to ensure that the log function is well defined. The parameter $\mu$ in this model is related to the infinitesimal shear modulus $\mu_0$ via:

$$\mu = \mu_0 \left[\frac{3-3N}{1-3N}\right]. \tag{2.2}$$

The parameter $N(> 1)$ is the number of (straight) links of the molecular chain, also known as the number of Kuhn segments. See the review by Puglisi and Saccomandi [19] for a microstructural interpretation of parameters in generalised neo-Hookean strain energy functions. We note that in the limit $N \to \infty$, the strain energy function $W$ of equation (2.1) reduces to the classical neo-Hookean model.

This model is a special case of the nonaffine network model first introduced by Davidson and Goulbourne [20]. Anssari-Benam and Bucchi [21] used this model to capture the deformation of the elastin isotropic matrix in heart valves. The model was later applied to the finite deformation of elastomers [18]. It belongs to the class of *limiting chain extensibility* models as the Gent model [22], and results in a similar Taylor series around $I_1 = 3$, see [18,23]. However, as shown by Horgan and Saccomandi [24], the Gent model is the simplest of the lowest order rational approximant in $I_1$, i.e. of the order [0/1]. In addition, the Gent model has no direct connection to a Padé approximant of any order of the inverse Langevin function [24]. In comparison, the model in equation (2.1) is a higher order rational approximant of $I_1$, of the order [1/1], and uses a Padé approximation of the [3/2] order of the inverse Langevin function, see [25] for a detailed analysis and demonstration on this point. In practice, these characteristics result in a more accurate fit of the proposed model to the experimental data compared with the Gent model, as previously demonstrated in, e.g., [18,26], for various elastomer datasets.

The model in equation (2.1) is of the generalised neo-Hookean type, as it is a function of $I_1$ only. Many studies, however, through either experimental observations [27-29] or theoretical analysis by way of deriving universal relationships [30-32], advocate the inclusion of an $I_2$ (the second invariant of the Cauchy-Green deformation tensors) term within the strain energy function $W$. See also [33] for a review of the central role of $I_2$ in modelling the nonlinear elasticity of rubber-like materials. Furthermore, as analysed at length by Destrade et al. [34], the inclusion of an $I_2$ term is also expedient from the perspective of compatibility with fourth-order weakly nonlinear elasticity. Moreover, the addition of an $I_2$-term will improve the modelling predictions in small-to-moderate deformation ranges; see [33] for direct comparisons and results involving the model (2.1). In this spirit, we consider the following $W(I_1, I_2)$ function based on the *parent* model in equation (2.1) as:

$$W = \mu N \left[\frac{1}{6N}(I_1 - 3) - \ln\left(\frac{I_1 - 3N}{3 - 3N}\right)\right] + C_2\left(\sqrt{\frac{I_2}{3}} - 1\right), \tag{2.3}$$

where $C_2$ is a positive material parameter. This particular $I_2$ term in equation (2.3) was first introduced by Carroll [35]. For a structural derivation of this term based on the concept of modelling the molecular chain entanglements as a topological tube constraint see [33,36]. Other customary $I_2$-term adjuncts, including a Mooney-Rivlin term $C_2(I_2 - 1)$ or a Gent-Thomas type term $C_2 \ln\left(\frac{I_2}{3}\right)$ [37,38] have been considered previously [33,39] and are not be pursued here. In essence, however, we note that all the foregoing adjunct $I_2$ terms produce a broadly similar contribution to the modelling results.

In summary, therefore, the set of invariant-based models from the proposed family of models which we wish to consider here are

$$\begin{cases} W = \mu N \left[\dfrac{1}{6N}(I_1 - 3) - \ln\left(\dfrac{I_1 - 3N}{3 - 3N}\right)\right], \\ \\ W = \mu N \left[\dfrac{1}{6N}(I_1 - 3) - \ln\left(\dfrac{I_1 - 3N}{3 - 3N}\right)\right] + C_2\left(\sqrt{\dfrac{I_2}{3}} - 1\right). \end{cases} \tag{2.4}$$

In the next section we demonstrate how these two models compare with each other and with the Ogden model in capturing the considered brain tissue deformation experimental datasets. For brevity, we refer to the models in equations $(2.4)_1$ and $(2.4)_2$ as '*Model I*' and '*Model II*', respectively. In §4 we introduce a third model, namely '*Model III*', which is expressed in terms of principal stretches instead of principal invariants (and, like the Ogden model, cannot be written easily in terms of $I_1$, $I_2$).

# 3. Application to experimental data

In this section the suitability and capability of the proposed models to capture Features (a)-(c) of the brain tissue mechanics are examined. As the first step, we start by applying the models in equation (2.4) to two comprehensive datasets due to Budday et al. [3,4] on compression, tension and simple shear deformations of the human brain (cortex) tissue. The collated datasets from those studies are presented in Tables A1 and A2 of Appendix A. These studies, and several other recent studies such as [5,6], suggest the application of the one-term Ogden model to the brain tissue deformation. Here we analyse the application of the Ogden model to these datasets and demonstrate the ensuing issues, while showing that no such issues arise on the use of our proposed models.

Starting with the preliminaries, the kinematics of the uniaxial and simple shear deformations may be characterised by the following deformation gradients:

$$\mathbf{F}_{uni} = \begin{bmatrix} \lambda & 0 & 0 \\ 0 & \lambda^{-0.5} & 0 \\ 0 & 0 & \lambda^{-0.5} \end{bmatrix}, \qquad \mathbf{F}_{ss} = \begin{bmatrix} 1 & \gamma & 0 \\ 0 & 1 & 0 \\ 0 & 0 & 1 \end{bmatrix}, \tag{3.1}$$

where the subscripts '*uni*' and '*ss*' denote the uniaxial and simple shear deformations, respectively, $\lambda$ is the amount of stretch and $\gamma$ the amount of shear. The Cauchy stress $\mathbf{T}$ for an isotropic incompressible material in finite deformation may be given in the following form:

$$\mathbf{T} = -p\mathbf{I} + 2W_1\mathbf{B} - 2W_2\mathbf{B}^{-1}, \tag{3.2}$$

where $\mathbf{B}$ $(= \mathbf{F}\mathbf{F}^{\mathrm{T}})$ is the left Cauchy-Green deformation tensor and $\mathbf{B}^{-1}$ is its inverse, $p$ is an arbitrary Lagrange multiplier enforcing the condition of incompressibility, $\mathbf{I}$ is the identity tensor, and $W_1$ and $W_2$ are the partial derivatives of the strain energy function $W$ with respect to the first and second principal invariants of $\mathbf{B}$, respectively, which are defined as:

$$I_1 = \mathrm{tr}(\mathbf{B}) = \lambda_1^2 + \lambda_2^2 + \lambda_3^2, \qquad I_2 = \mathrm{tr}(\mathbf{B}^{-1}) = \lambda_1^{-2} + \lambda_2^{-2} + \lambda_3^{-2}, \tag{3.3}$$

with the third invariant $I_3 = \det(\mathbf{B}) = \lambda_1^2\lambda_2^2\lambda_3^2 = 1$ due to incompressibility. On using the models given by equation (2.4) and the deformation gradients by equation (3.1) in equation (3.2) we obtain:

$$T_{uni} = \frac{\mu}{3}\left(\frac{\lambda^2 + 2\lambda^{-1} - 9N}{\lambda^2 + 2\lambda^{-1} - 3N}\right)\left(\lambda^2 - \frac{1}{\lambda}\right) + \frac{C_2}{\sqrt{3(\lambda^{-2} + 2\lambda)}}\left(\lambda - \frac{1}{\lambda^2}\right), \qquad (3.4)$$

for the uniaxial (compression-tension) deformation of *Model II*, taking $C_2 = 0$ for *Model I*. Note that in this case $I_1 = \lambda^2 + 2\lambda^{-1}$ and $I_2 = \lambda^{-2} + 2\lambda$. Similarly, for simple shear,

$$T_{ss} = \left[\frac{\mu}{3}\left(\frac{3 + \gamma^2 - 9N}{3 + \gamma^2 - 3N}\right) + \frac{C_2}{\sqrt{3(3 + \gamma^2)}}\right]\gamma, \qquad (3.5)$$

where $I_1 = I_2 = 3 + \gamma^2$, for *Model II*, and taking $C_2 = 0$ for *Model I*.

To obtain the optimised fit, these equations (for each model) were simultaneously fitted to the uniaxial and simple shear deformation datasets where the best fit is achieved by minimising the residual sum of squares (RSS) function defined as: $\text{RSS} = \sum_i \left(\frac{T^{model} - T^{experiment}}{T^{experiment}}\right)_i^2$, where $i$ is the number of data points. This method minimises the relative error, as opposed to the absolute error. The importance of this approach has been highlighted by Destrade et al. [34]. Here the minimisation was performed via an in-house developed code in MATLAB®.

We now proceed with presenting the fitting results provided by the proposed models, starting with the data in [3]. First, we consider the compression and shear datasets, as perhaps the most relevant loading modalities for brain deformation applications since they directly relate to common injuries such as the traumatic brain and diffuse axonal injuries. The fitting results for *Model I* and *Model II*, i.e., using equations (3.4) and (3.5), are presented in figure 1. Panels in the right-hand side of the figure present the relative error (%), defined as $\left|\frac{T^{model} - T^{experiment}}{T^{experiment}}\right| \times 100$. The values of the model parameters are summarised in table 1. Note that here the optimisation procedure yields a very small value for $C_2$, so that both models produce almost identical fitting results with no discernible deviation. As the plots and the $R^2$ values indicate, both models provide an excellent fit to the dataset.

These results provide a first verification of the capability of *Models I* and *II* to capture Features (a) and (c) of brain tissue mechanics proposed by studies [3,4,13,14], as listed in §1. However, to investigate Feature (b) in more depth it is necessary to explore the performance of the proposed models when tension data is also included. Figure 2a illustrates the application of *Models I* and *II* to compression-tension and simple shear deformation data (again, the datasets are from [3]). Table 2 summarises the values of the model parameters obtained in this fit. Note that for *Model II*, $C_2$ is very small, so that *Models I* and *II* produce a similar fit again.

As the $R^2$ values in table 2 indicate, the models provide a fairly favourable agreement with the experimental data. However, compared with the case where only uniaxial compression was considered (figure 1), the model predictions in figure 2a show a higher deviation from the experimental data. This perceived deviation is somewhat reconciled if the typical error bars are also taken into account (as reported in, e.g., [1,4]). Nevertheless, the models encounter difficulty in capturing the full asymmetry of the compression-tension behaviour.

To this end, Budday et al. [3] advocate the application of the one-term Ogden model to this dataset, which we use for comparison in the following. This celebrated strain-energy function is of the form [40]:

$$W_{Ogden} = \frac{2\mu}{\alpha^2} \left(\lambda_1^\alpha + \lambda_2^\alpha + \lambda_3^\alpha - 3\right), \tag{3.7}$$

where $\mu$ is the infinitesimal shear modulus (in Pa) and $\alpha$ is a real, non-dimensional material parameter. The components of the corresponding Cauchy stress for uniaxial tension and simple shear deformations are:

$$\begin{cases} T_{Uni} = \dfrac{2\mu}{\alpha} \left(\lambda_1^\alpha - \lambda_1^{-0.5\alpha}\right), \\ T_{ss} = \dfrac{\mu}{\alpha\sqrt{1+\dfrac{\gamma^2}{4}}} \left[\left(\dfrac{\gamma}{2} + \sqrt{1+\dfrac{\gamma^2}{4}}\right)^\alpha - \left(-\dfrac{\gamma}{2} + \sqrt{1+\dfrac{\gamma^2}{4}}\right)^\alpha\right], \end{cases} \tag{3.8}$$

respectively.

When the relationships in equation (3.8) are simultaneously fitted to the data from [3] using the same fitting process described earlier in this section, the best fit is obtained with $\mu = 1.04$ kPa and $\alpha = -17.33$, producing $R^2$ values of 0.95 and 0.88 for the uniaxial compression-tension and simple shear deformation datasets, respectively. The fitting results are illustrated in figure 2b.

We observe that while the Ogden model undoubtedly better captures the asymmetry in the compression-tension behaviour, the relative errors for the fits provided to this dataset are broadly comparable for *Models I* and *II* and for the Ogden model – see the relative error panel in figure 2c. However, beyond the marginal improvements that one model may provide over the other, most of all we note the non-convexity of the iso-energy plots in the plane of the principal stretches $\lambda_1$ and $\lambda_2$ produced by fitting the one-term Ogden model to the considered dataset, as presented in figure 3a. By contrast, *Models I* and *II* in equation (2.4) are *a priori* convex over their defined domain of deformation – see [18,33] for further discussion and analysis. Figure 3b illustrates the iso-energy plots of *Model I* for the same dataset, as a representative example of the proposed models. Model parameters are those given in table 2.

Applying the one-term Ogden model to the considered brain deformation dataset has an additional undesirable effect. By inspecting the $T_{Uni} - \lambda$ plot of the Ogden model in figure 2b it may be qualitatively observed that the provided fit does not correctly capture the concavity of the experimental data in the tensile range (i.e., $\lambda > 1$). This can be illustrated quantitatively by computing the second derivative of $T_{Uni}$ with respect to $\lambda$, which on using equation $(3.8)_1$ and the characterised values of $\mu = 1.04$ kPa and $\alpha = -17.33$ from the fit, is:

$$\frac{\partial^2 T_{Uni}}{\partial \lambda^2} = \frac{2(\alpha-1)\mu\lambda^{\frac{3\alpha}{2}} - \left(1+\frac{\alpha}{2}\right)\mu}{\lambda^{2+\frac{\alpha}{2}}} = \frac{-38.1264\lambda^{-25.995} + 7.9716}{\lambda^{-6.665}}. \tag{3.9}$$

Hence, we see from equation (3.9) that for the tensile range $1 < \lambda < 1.061$ (first 6% of extension), the one-term Ogden model provides a negative concavity, whereas the experimental data exhibits a positive concavity. No such discrepancy arises from the application of the proposed models in equation (2.4) to this dataset.

As another example, we highlight the study of Budday et al. [4], which provides a comprehensive set of uniaxial compression-tension and simple shear deformation datasets for various anatomical regions of the human brain, including the cortex. The collated dataset from that study pertaining to the deformation of the cortex is presented in table A2. Figure 4a demonstrates the fitting results using *Models I* and *II*, and table 3 summarises the model parameters. Similar to the previous dataset, the proposed set of models provides fairly favourable fits to the data; see the $R^2$ values in table 3. Again, while the models do not seem to have an inherent shortcoming to predict the compression-tension data, it is evident that capturing the asymmetry of the compression-tension behaviour can be improved.

In a similar fashion to [3], Budday et al. [4] also consider the one-term Ogden model as optimal for capturing the dataset and the asymmetry in the compression-tension behaviour. By fitting the stress-stretch relationships in equation (3.8) to the dataset in [4], the best fit provided by the one-term Ogden model is obtained with $\mu = 1.46$ kPa and $\alpha = -19.12$, resulting in $R^2$ values of 0.99 for both uniaxial compression-tension and simple shear deformation datasets. See figure 4b for the results. By comparing the relative error plots presented in figure 4c, it is evident that the one-term Ogden model provides an improved fit to the compression-tension data, while *Models I* and *II* better capture the shear deformation.

However, once more, irrespective of the fitting advantages of one model over the other, the problem of non-convexity using the one-term Ogden model manifests itself again. By plotting the iso-energy graphs in the plane of the principal stretches $\lambda_1$ and $\lambda_2$ produced by the fit that the one-term Ogden model provides to the considered dataset, with $\mu = 1.46$ kPa and $\alpha = -19.12$, we find that the ensuing strain energy function is non-convex (figure 5a). By contrast, the strain energies of *Models I* and *II* (with the material parameters in table 3) remain convex over their defined domain of deformation (figure 5b).

In addition, the same problem in capturing the incorrect concavity of the experimental compression-tension data within a range of the tensile deformation domain, as per the previous dataset [3], is present in the predictions of the one-term Ogden model for this dataset too. Equation (3.9) allows to analyse this observation mathematically. With $\mu = 1.46$ kPa and $\alpha = -19.12$ for this data set we find that for the tensile range $1 < \lambda < 1.056$, the Ogden model provides a negative concavity, whereas the experimental data exhibit a positive concavity. No such issues arise from the application of *Models I* and *II* in equation (2.4) to this dataset.

# 4. A generalisation of the proposed models

We now note a study by Mihai et al. [41] reporting and modelling experimental data on human brain tissue. Their paper proposed a special three-term Ogden model by the addition of a Mooney-Rivlin term, with the following final form *mutatis mutandis* as:

$$W = \frac{\mu_1}{\alpha}\left(\lambda_1^{\alpha} + \lambda_2^{\alpha} + \lambda_3^{\alpha} - 3\right) + \mu_2\left(\lambda_1^{2} + \lambda_2^{2} + \lambda_3^{2} - 3\right) + \mu_3\left(\lambda_1^{-2} + \lambda_2^{-2} + \lambda_3^{-2} - 3\right), \qquad (4.1)$$

where $\mu_1$, $\mu_2$ and $\mu_3$ are stress-like constants, and $\alpha$ is a nondimensional stiffening constant. The motivation behind the development of this model was to achieve improved fits compared with the one-term Ogden model while avoiding the excessive number of model parameters that may otherwise arise due to the use of the general higher-term Ogden models [41]. As demonstrated in [41], the model in equation (4.1) provides the sought improved fits with only four parameters $(\mu_1, \mu_2, \mu_3, \alpha)$, whereas the equivalent Ogden model would have six material parameters $(\mu_1, \mu_2, \mu_3, \alpha_1,\ \alpha_2,\ \alpha_3)$. The calibration of the model with the considered experimental dataset therein was achieved with $\mu_1 = 0.0653$ kPa, $\mu_2 = -0.1151$ kPa, $\mu_3 = -0.9921$ kPa, and $\alpha = 14.3626$. However, we note that the non-convexity of the strain energy function still persists, as plotted in figure 6a.

Incidentally, when, in the spirit of Mihai et al [41], we add the one-term Ogden model in equation (3.7) to, say, *Model I* in equation (2.4)$_1$, as follows:

$$\mathcal{W} = \frac{2\mu_1}{\alpha^2}\left(\lambda_1^\alpha + \lambda_2^\alpha + \lambda_3^\alpha - 3\right) + \mu_2 N\left[\frac{1}{6N}(I_1 - 3) - \ln\left(\frac{I_1 - 3N}{3 - 3N}\right)\right], \tag{4.2}$$

the ensuing strain energy function $\mathcal{W}$ provides much improved fits to the uniaxial compression-tension and simple shear datasets compared with that originally provided by *Model I* only, while remaining convex over the domain of deformation, see figure 6b. The plots in figure 7 demonstrate the fitting results for the dataset due to Budday et al. [4] with the *modified Model I* in equation (4.2) as a representative example. The best fit was obtained with $\mu_1 = 0.9781$ kPa, $\alpha = -23.4610$, $\mu_2 = 0.3672$ kPa and $N = 4.4578$, and $R^2$ values in excess of 0.99.

Treloar [42] stipulated that Rivlin's representation of a strain energy function $W$ in invariant form was so considered to keep $W$ as an even-powered function of the stretch ratios, hence $W$ would always be positive. By noting that the principal stretches $\lambda_i$ are by definition positive, Ogden removed the overly prescriptive restriction of the even-powered $\lambda_i$ terms in his formulation, to let the powers $\alpha_i$ be real-valued [40]. However, while the Ogden model is an extension to the functional form of the Valanis-Landel [43] representation, it is not the most general form of $W$ as a function of $\lambda_i$, since $\lambda_i$ terms in the Ogden model are also in a separable form. There is no mathematical or physical necessity for a strain energy function to be a separable form of $\lambda_i$, and the most general case is achieved via non-separable forms. For a detailed discussion on this point see [42].

In this spirit, a generalisation of the models in equation (2.4) may be achieved by considering the following, non-separable, representation:

$$W = \mu N\left[\frac{1}{6N}\left(\lambda_1^\alpha + \lambda_2^\alpha + \lambda_3^\alpha - 3\right) - \ln\left(\frac{\lambda_1^\alpha + \lambda_2^\alpha + \lambda_3^\alpha - 3N}{3 - 3N}\right)\right], \tag{4.3}$$

where $\alpha$ is real-valued. We refer to this model as *Model III* of the family of the proposed models herein. Setting $\alpha = 2$ recovers *Model I* given by equation (2.1). We note that *Model III* is a hybrid of structural basis and phenomenological considerations. We find from equation (4.3) the expressions for the Cauchy stress in uniaxial and simple shear deformations as:

$$\begin{cases} T_{Uni} = \dfrac{\mu\alpha\left(\lambda^\alpha + 2\lambda^{-0.5\alpha} - 9N\right)}{6\left(\lambda^\alpha + 2\lambda^{-0.5\alpha} - 3N\right)}\left(\lambda^\alpha - \lambda^{-0.5\alpha}\right), \\ T_{ss} = \dfrac{\mu\alpha\left(\lambda^\alpha + 2\lambda^{-0.5\alpha} - 9N\right)}{6\sqrt{\gamma^2+4}\left(\lambda^\alpha + 2\lambda^{-0.5\alpha} - 3N\right)}\left(\lambda^\alpha - \lambda^{-0.5\alpha}\right), \end{cases} \tag{4.4}$$

where we recall that the principal stretch $\lambda$ in simple shear is given in terms of the amount of shear $\gamma$ as [16]:

$$\lambda = \frac{\gamma}{2} + \sqrt{1 + \frac{\gamma^2}{4}}\ . \tag{4.5}$$

Note that on specialising equation (4.4)$_2$ to the linear theory of elasticity, the infinitesimal shear modulus $\mu_0$ is found as:

$$\mu_0 = \frac{\mu\alpha^2(1 - 3N)}{4(3 - 3N)}\ . \tag{4.6}$$

By simultaneously fitting of the relationships in equation (4.4) with the compression-tension and shear deformation datasets of human brain cortex tissue specimens due to Budday et al. [3,4], most favourable fits are obtained compared with *Models I* and *II* in equation (2.4). The fitting results are illustrated in figure 8. Table 4 summarises the model parameters. By comparing the plots in figure 8 with those provided by the (one-term) Ogden model to the same datasets shown in figures 2 and 4, we observe that *Model III* in equation (4.3) provides improved fits to the Ogden model, particularly in capturing the asymmetry of the deformation in compression and tension, while remaining convex; see the insets in figure 8.

# 5. Summary and concluding remarks

The aim of this manuscript was to make a case for the application of our proposed family of models, namely *Models I*, *II* and *III*, given by equations (2.4) and (4.3), to the mechanics of brain tissue. To this end, the analyses presented herein demonstrate that these models are inherently capable of capturing the key features of the mechanical behaviour of brain tissue according to Budday and collaborators [3,4,14] and Mihai et al. [15]; i.e., Features (a) to (c) listed in §1.

While the proposed models do not equally provide as favourable fits to all the datasets, they appear to perform better than most of the existing models in the literature. The only model capable of providing an improved fit to *Models I* and *II* seems to be the one-term Ogden model; however, the analyses presented in §§3 and 4 identify some problematic effects that arise from the application of the Ogden model to the brain tissue deformation datasets, such as the non-convexity of the strain energy function and the misplacement of the direction of concavity in the stress-stretch curves over a range of the deformation domain. By contrast, the family of models presented here proved impervious to such shortcomings. Pending further investigation, the epilogue analysis in §4 indicates that the phenomenological extension of the proposed models into a functional representation embodying the principal stretches $\lambda_i$, i.e., *Model III* as given by equation (4.3), provides an optimal solution where the model appears to capture the experimental data most favourably, while maintaining convexity.

We conclude with some comments on the first part of Feature (a): "The shear response is highly nonlinear". As noted by Budday and collaborators [1,12], this strong nonlinearity could be a consequence of testing protocol chosen in [3,4,14,15], where cubic samples were used. With that geometry it becomes impossible to ensure homogeneity of the shear deformation, and then the analytical expression for $T$ is not adequate to predict the data, as the nonlinearity is exacerbated by inhomogeneous effects, see Finite Element simulations [44] in figure 9. The high (absolute) values of the Ogden parameter $\alpha$ lead to "unrealistic high shear stresses" [1] outside the data range. Budday and collaborators [1,12] recommend an inverse Finite Element analysis to model the inhomogeneous response of a cubic sample and deduce the actual value of $\alpha$. Another route is to follow the recommendations of the standard protocols (e.g., that of the British Standards [45]), which recommend that the width of the sample be four times its height. With that precaution, the shear response is almost linear [28] or slightly nonlinear [6], indicating that the value of $|\alpha|$ for brain matter should be close to 2 (also confirmed by Balbi et al. [29] with torsion experiments on brain matter, and by Budday et al. [1], who find that the surface morphology due to brain development is most realistic in computational simulations when $\alpha = 2$). With that correction, the problems of non-convexity and wrong curvature of the tensile stress-stretch response raised in this paper for the Ogden model will probably disappear. At any rate, our proposed models, *Models I*, *II* and *III*, remain impervious to these shortcomings.

# Additional Information

**Data Accessibility**
The datasets used in this study have been provided in a tabulated format in Appendix A, tables A1 and A2.

**Authors' Contributions**
AAB: Developed the idea, modelling formulations and the results presented in this work, and wrote an initial version of the manuscript; MD: Checked the calculations, derivations and the results, and revised, rewrote and corrected various portions of the article; GS: Critically reviewed the manuscript and rewrote various parts of the article. All authors have approved the final version of the manuscript and are responsible for all aspects of the work.

**Competing Interests**
The authors declare that they have no competing interests.

**Funding Statement**
The research of MD was supported by a *111 Project for International Collaboration* (Chinese Government, PR China) No. B21034 and by a grant from the *Seagull Program* (Zhejiang Province, PR China). GS is supported by the *Istituto Nazionale di Alta Matematica* (INdAM, Italy), *Fondi di Ricerca di Base UNIPG* (Italy), the MIUR-PRIN project 2017KL4EF3 and Istituto Nazionale di Fisica Nucleare through its IS 'Mathematical Methods in Non-Linear Physics' (Italy).

# Appendix A: Tabulated experimental data

**Table A1** – Budday et al. data [3].

| Compression-Tension | | Simple Shear | |
|---|---|---|---|
| $\lambda$ | $T$ [kPa] | $\gamma$ | $T$ [kPa] |
| 0.90 | -0.58 | -0.20 | -0.40 |
| 0.91 | -0.44 | -0.17 | -0.30 |
| 0.925 | -0.34 | -0.15 | -0.23 |
| 0.94 | -0.25 | -0.125 | -0.17 |
| 0.95 | -0.18 | -0.10 | -0.13 |
| 0.96 | -0.12 | -0.07 | -0.09 |
| 0.975 | -0.07 | -0.05 | -0.06 |
| 0.99 | -0.03 | -0.02 | -0.03 |
| 1.00 | 0 | 0 | 0 |
| 1.01 | 0.015 | 0.02 | 0.03 |
| 1.02 | 0.04 | 0.05 | 0.07 |
| 1.035 | 0.065 | 0.07 | 0.10 |
| 1.05 | 0.09 | 0.10 | 0.15 |
| 1.06 | 0.11 | 0.12 | 0.22 |
| 1.07 | 0.15 | 0.15 | 0.30 |
| 1.085 | 0.18 | 0.17 | 0.41 |
| 1.10 | 0.25 | 0.20 | 0.58 |

**Table A2** – Budday et al. data [4].

| Compression-Tension | | Simple Shear | |
|---|---|---|---|
| $\lambda$ | $T$ [kPa] | $\gamma$ | $T$ [kPa] |
| 0.90 | -1.06 | -0.20 | -0.55 |
| 0.91 | -0.82 | -0.175 | -0.38 |
| 0.92 | -0.62 | -0.15 | -0.27 |
| 0.94 | -0.46 | -0.125 | -0.19 |
| 0.95 | -0.33 | -0.1 | -0.14 |
| 0.96 | -0.225 | -0.075 | -0.1 |
| 0.975 | -0.135 | -0.05 | -0.06 |
| 0.99 | -0.06 | -0.02 | -0.04 |
| 1.00 | 0 | 0 | 0 |
| 1.01 | 0.04 | 0.02 | 0.03 |
| 1.025 | 0.08 | 0.05 | 0.06 |
| 1.04 | 0.12 | 0.07 | 0.1 |
| 1.05 | 0.15 | 0.10 | 0.14 |
| 1.06 | 0.19 | 0.12 | 0.19 |
| 1.075 | 0.25 | 0.15 | 0.27 |
| 1.09 | 0.34 | 0.17 | 0.37 |
| 1.10 | 0.46 | 0.20 | 0.54 |

# Tables and captions

**Table 1** – Model parameters for the compression and simple shear deformation datasets of human brain cortex reported in Budday et al. [3] using the model in equation (2.4). The $R^2$ values are presented in (compression) – (simple shear) order.

| | $\mu$ [kPa] | $N$ [-] | $C_2$ [Pa] | $R^2$ (comp. – ss) |
|---|---|---|---|---|
| *Model I* | 0.022 | 1.02 | --- | 0.99 – 0.97 |
| *Model II* | 0.040 | 1.02 | $8.15 \times 10^{-3}$ | 0.99 – 0.97 |

**Table 2** – Model parameters for the compression-tension and simple shear deformation datasets of human brain cortex reported in Budday et al. [3]. The $R^2$ values are presented in the (uniaxial) – (simple shear) order.

| | $\mu$ [kPa] | $N$ [-] | $C_2$ [Pa] | $R^2$ (uni. – ss) |
|---|---|---|---|---|
| *Model I* | 0.022 | 1.02 | --- | 0.88 – 0.95 |
| *Model II* | 0.022 | 1.02 | $3.57 \times 10^{-3}$ | 0.88 – 0.95 |

**Table 3** – Model parameters for the uniaxial compression-tension and simple shear deformation datasets of human brain cortex reported in Budday et al. [4]. The $R^2$ values are presented in (uniaxial) – (shear) deformation order.

| | $\mu$ [kPa] | $N$ [-] | $C_2$ [Pa] | $R^2$ (uni. – ss) |
|---|---|---|---|---|
| *Model I* | 0.05 | 1.025 | --- | 0.87 – 0.98 |
| *Model II* | 0.01 | 1.017 | 4.00 | 0.90 – 0.99 |

**Table 4** – Model parameters for the uniaxial compression-tension and simple shear deformation datasets of human brain cortex due to Budday et al. [3,4] using *Model III* in equation (4.3). The $R^2$ values are presented in (uniaxial) – (shear) deformation order.

| | $\mu$ [kPa] | $N$ [-] | $\alpha$ [-] | $R^2$ (uni. – ss) |
|---|---|---|---|---|
| Budday et al. [3] data | 0.02 | 1.16 | -5.46 | 0.95 – 0.94 |
| Budday et al. [4] data | 0.02 | 5.50 | -16.00 | 0.99 – 0.99 |

# Figures and captions

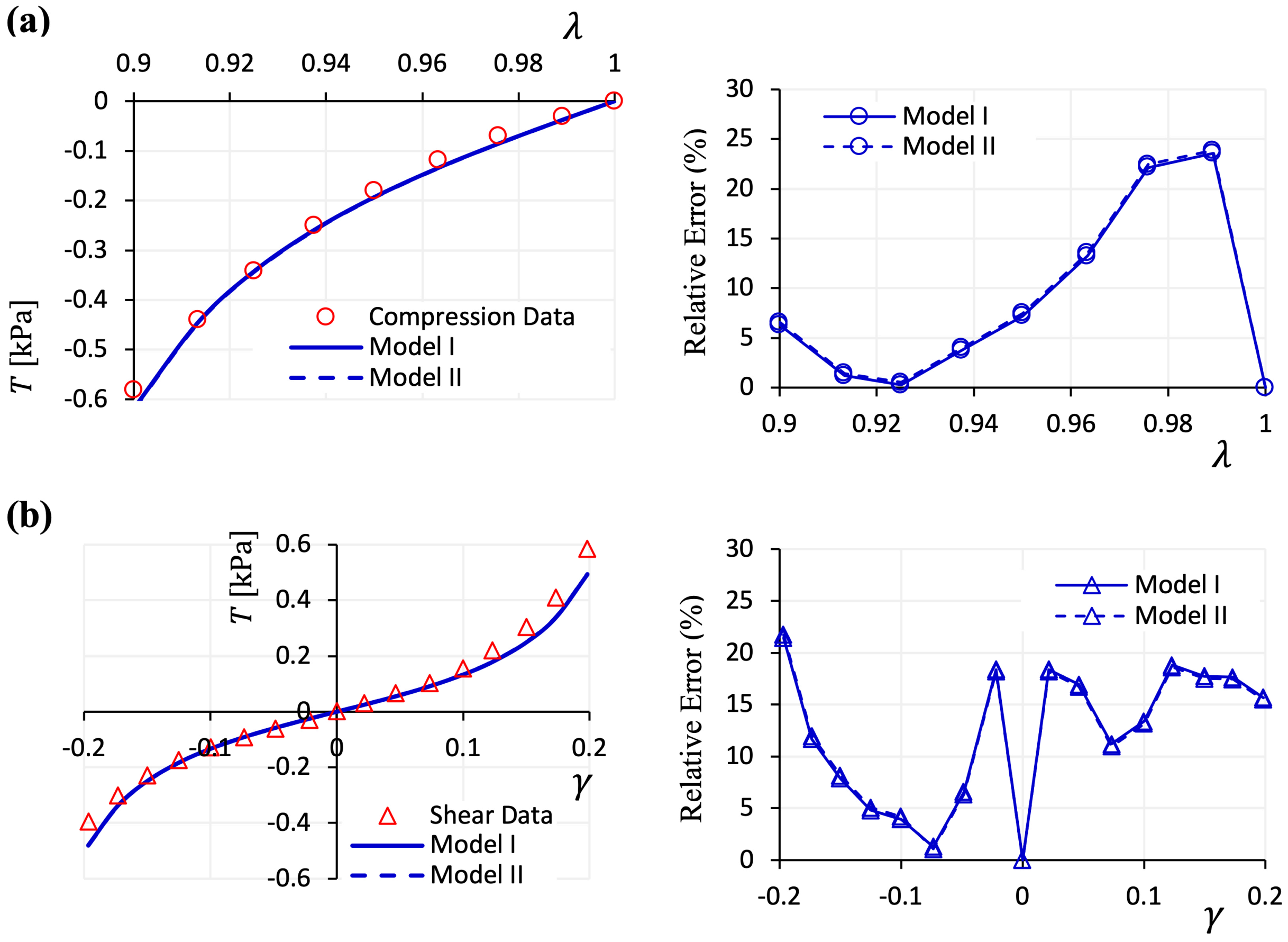

**Figure 1** – Modelling results for the (a) compression and (b) simple shear deformations of human brain tissue using *Models I* and *II* in equation (2.4). The right-hand side panels demonstrate the relative errors. Both models produce almost an identical fit, because the optimisation procedure gives a very small value for $C_2$ in *Model II*. Dataset for human brain cortex due to Budday et al. [3].

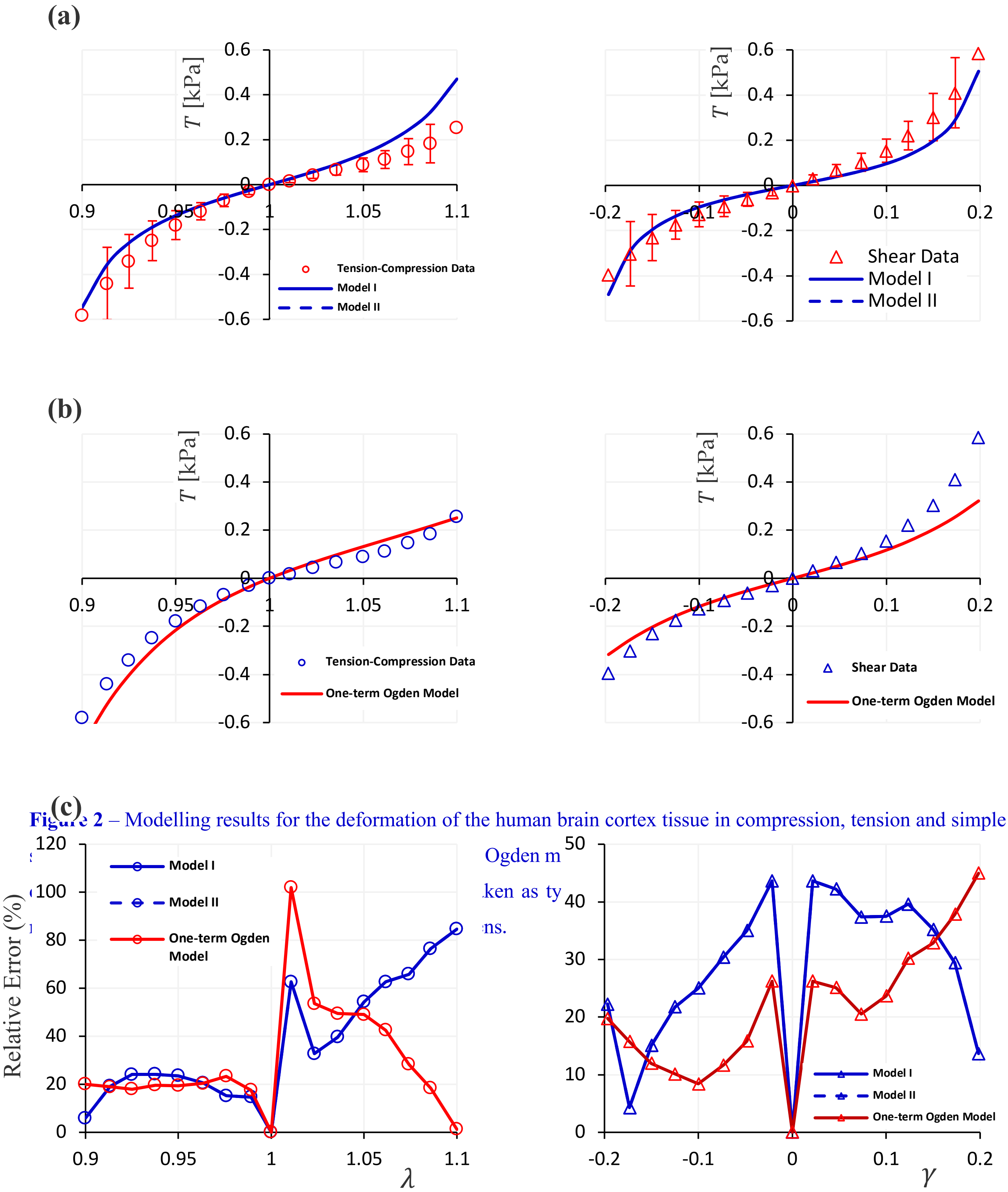

**Figure 2** – Modelling results for the deformation of the human brain cortex tissue in compression, tension and simple ... Ogden m... ken as ty... ns.

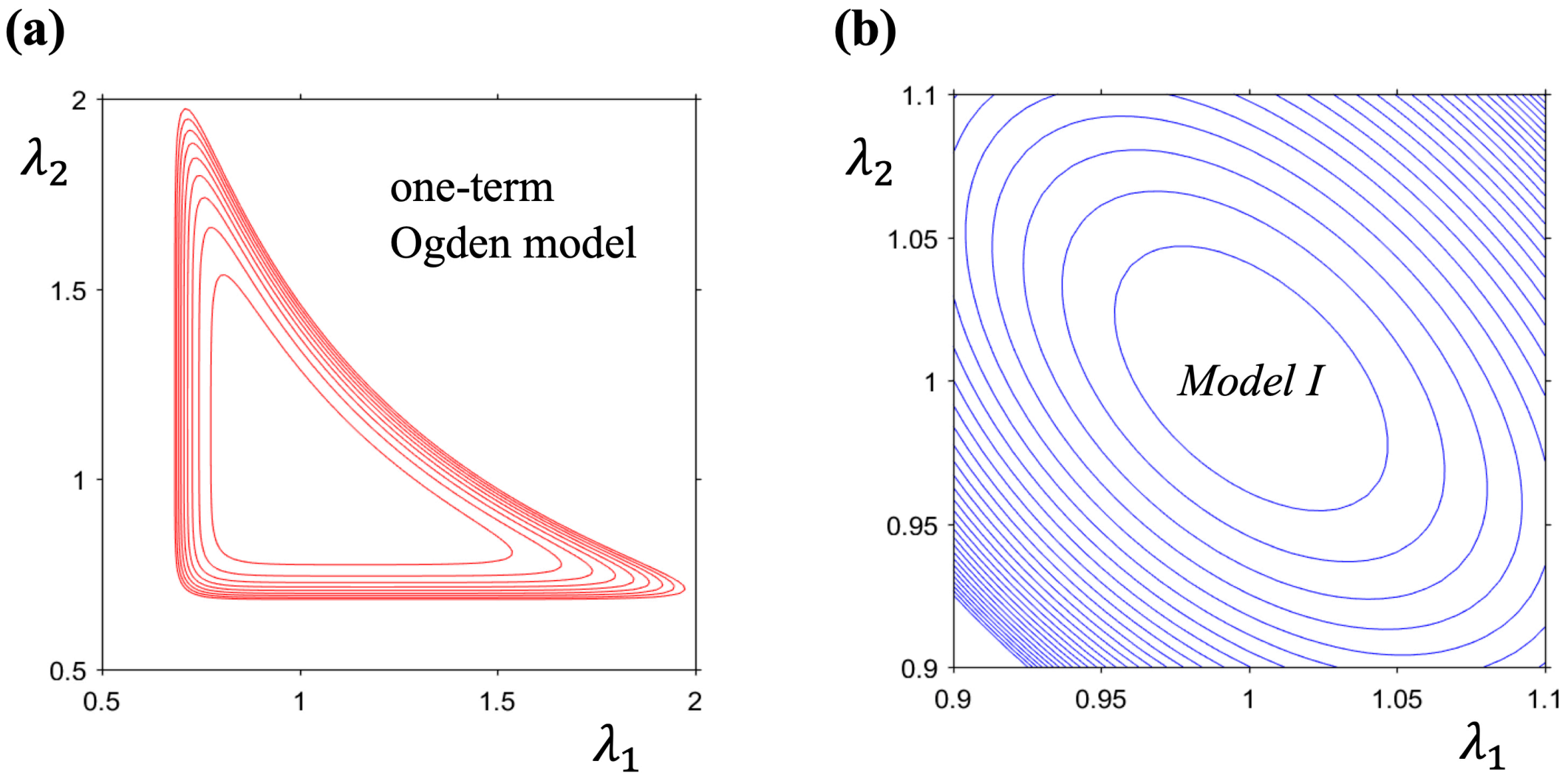

**Figure 3** – The iso-energy plots in the plane of principal stretches $(\lambda_1, \lambda_2)$ for: (a) the one-term Ogden model; and (b) the proposed *Model I*, for the uniaxial compression-tension and simple shear datasets of Budday et al. [3]. The non-convexity of the Ogden model is evident, while the proposed model remains convex *a priori* over the domain of deformation.

(a)

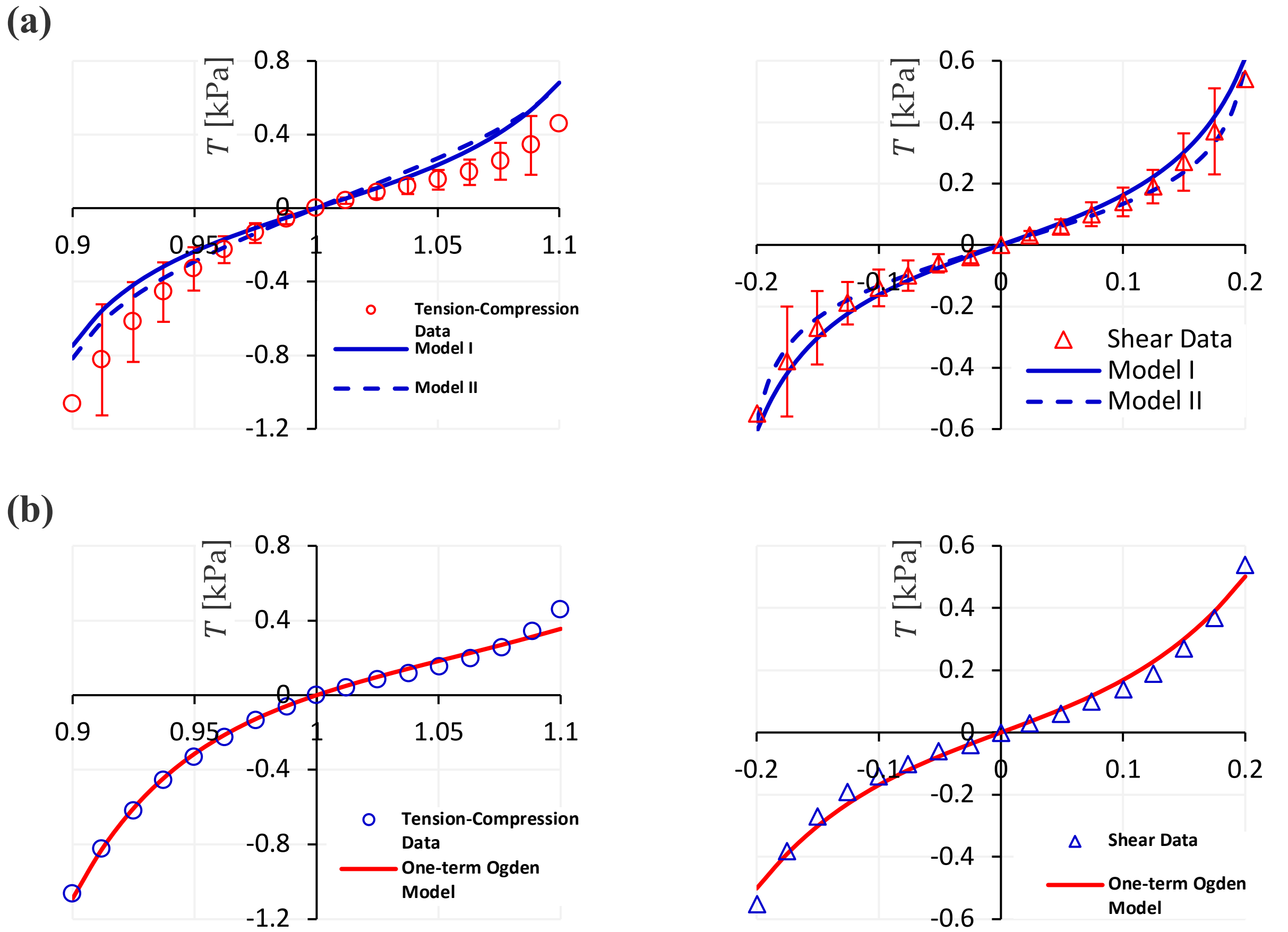

**Figu**[illegible] – Modelling results for the deformation of the human brain cortex tissue in compression, tension and simple shear: (a) using *Models I* and *II*; (b) using the one-term Ogden model; and (c) the ensuing relative errors. The dataset is due

(c)

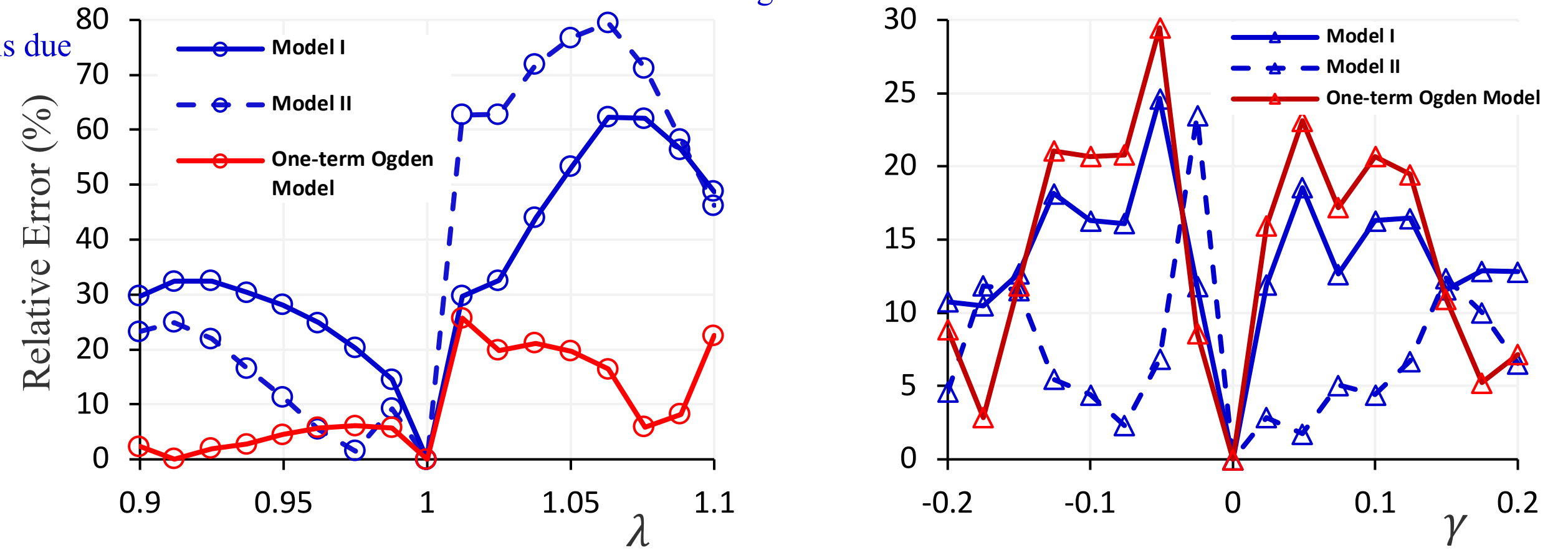

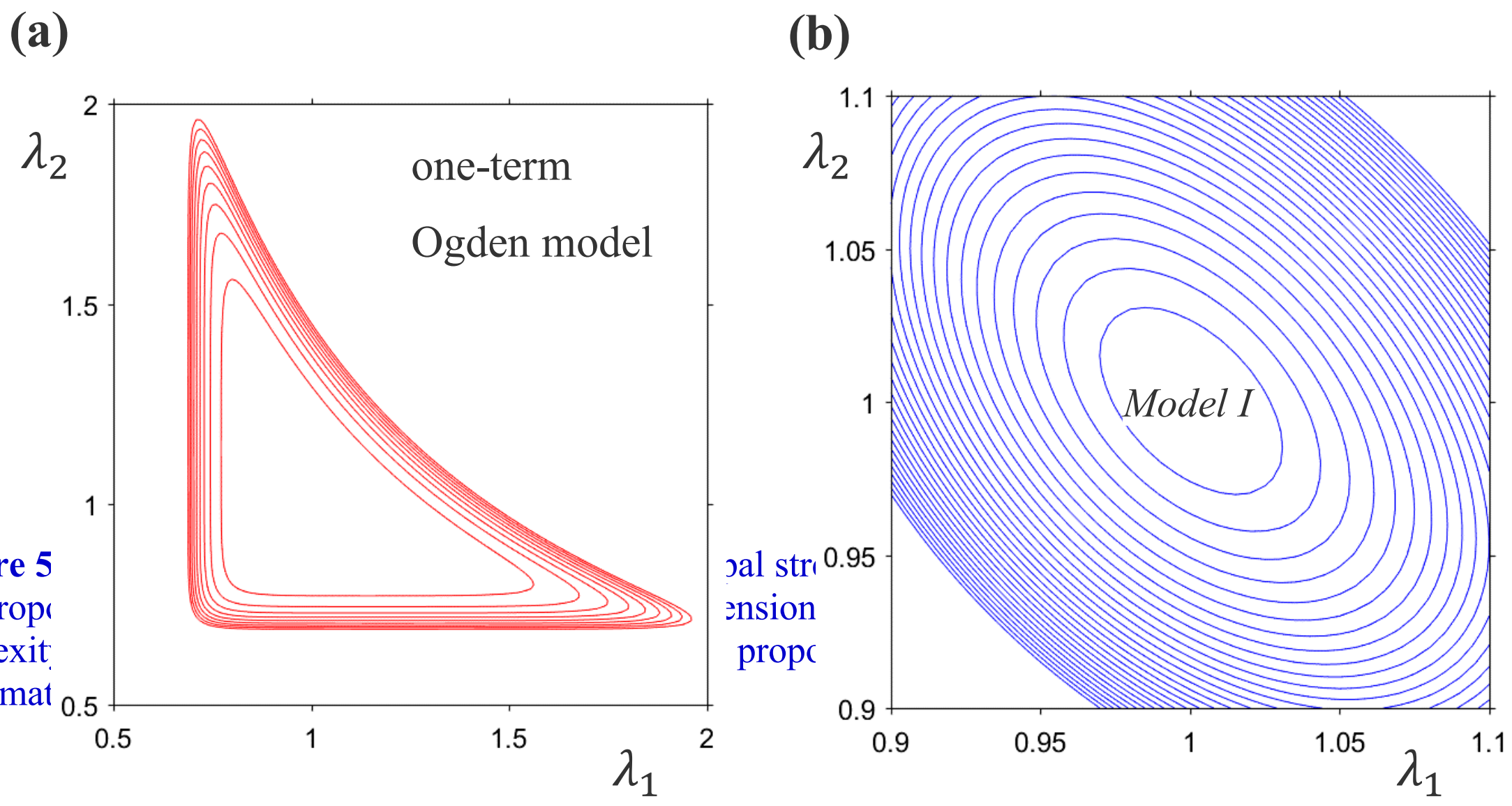

**Figure 5** … bal str… …del; and (b)
the prop… …ension… …]. The non-
convexit… propc… …domain of
deformat…

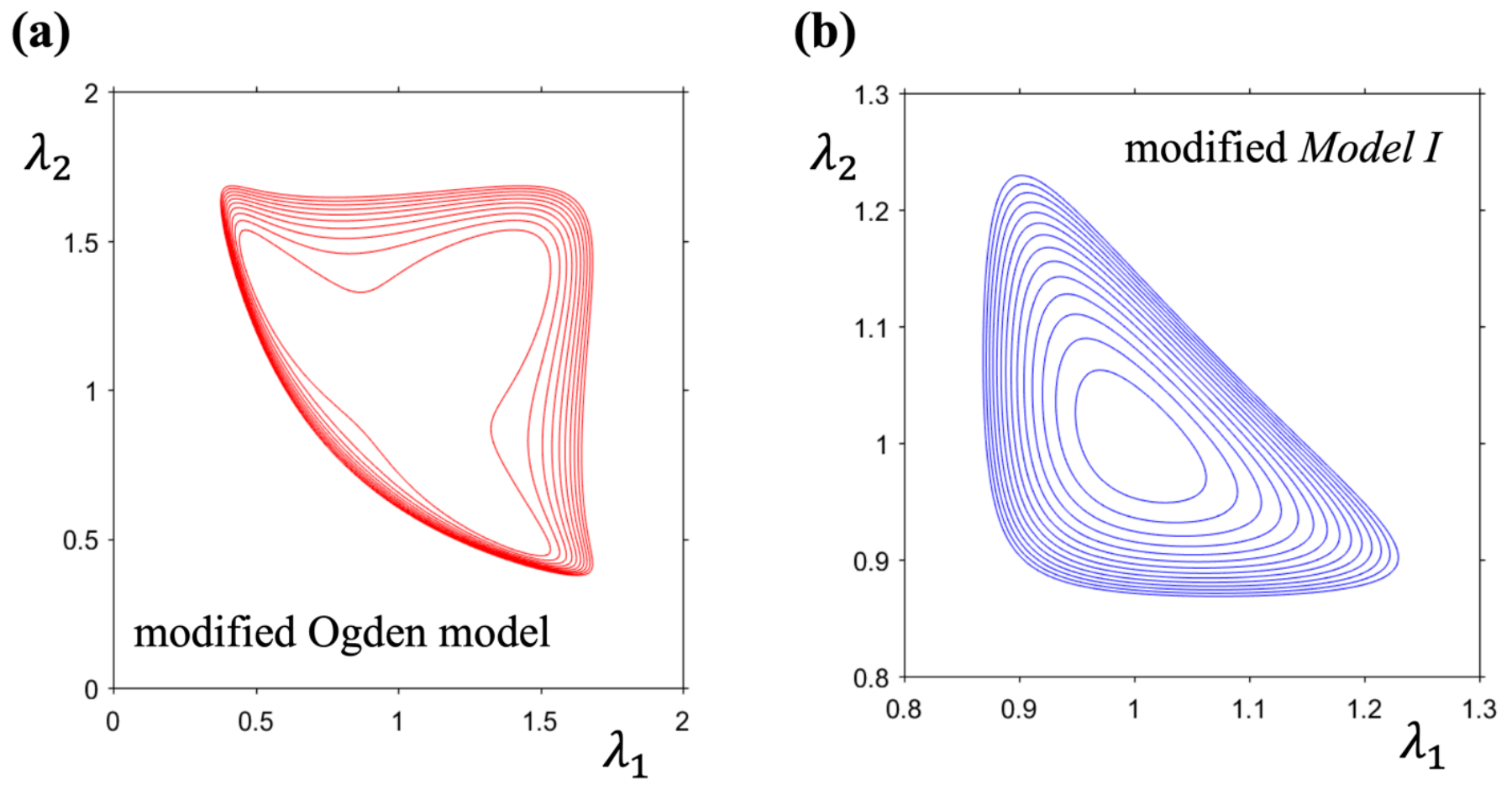

**Figure 6** – The iso-energy plots of (a) the modified one-term Ogden model in equation (4.1) presented by Mihai et al [41] and (b) the *modified Model I* in equation (4.2).

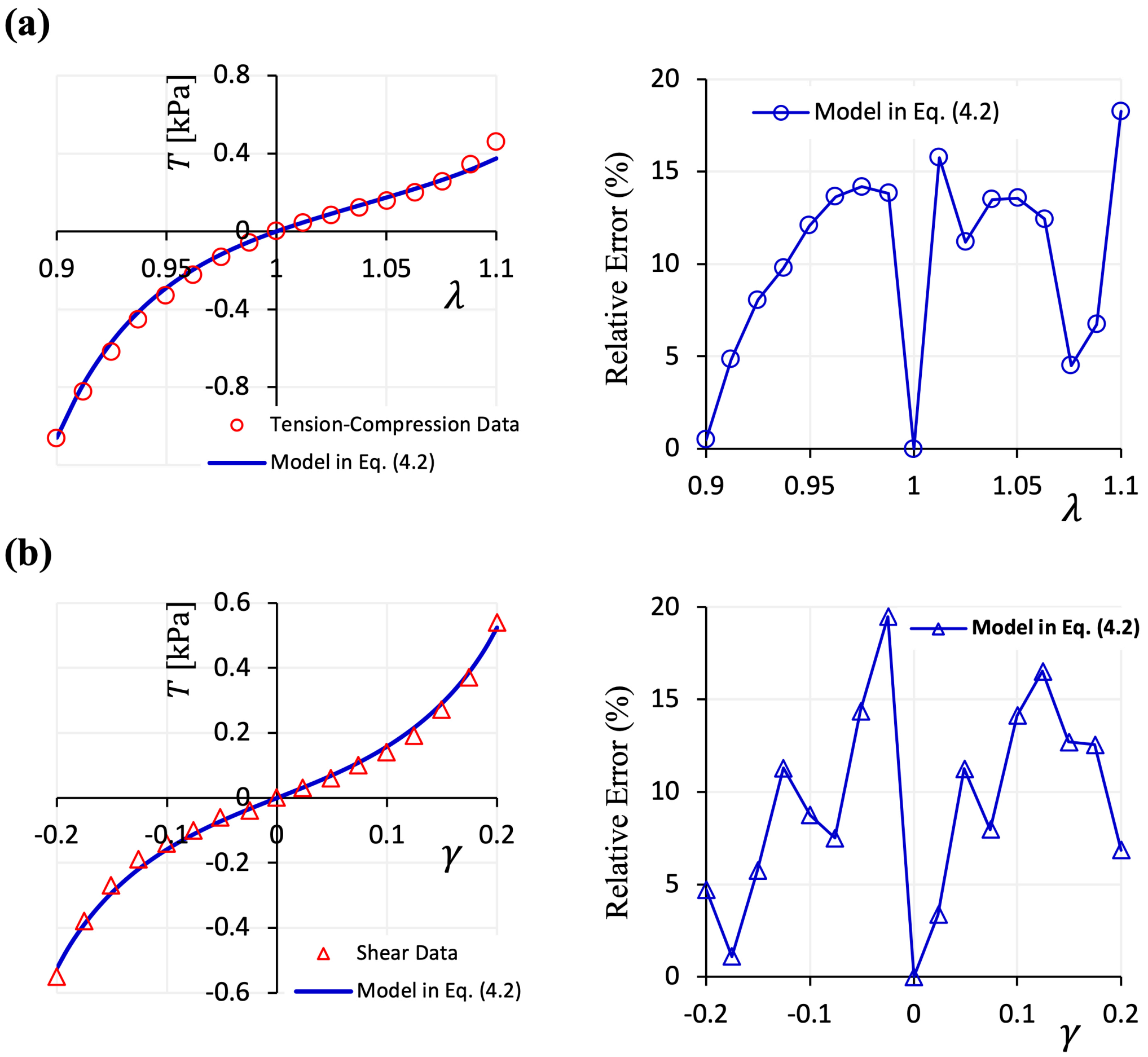

**Figure 7** – The fitting results using the strain energy function in equation (4.2) for the *modified Model I* to the dataset of Budday et al. [4]: (a) the compression-tension data; (b) the simple shear data. Model parameters for the obtained fits are: $\mu_1 = 0.9781$ kPa, $\alpha = -23.4610$, $\mu_2 = 0.3672$ kPa and $N = 4.4578$. Note the significant improvement in the quality of the fits compared with the plots in figure 4a using *Model I* only.

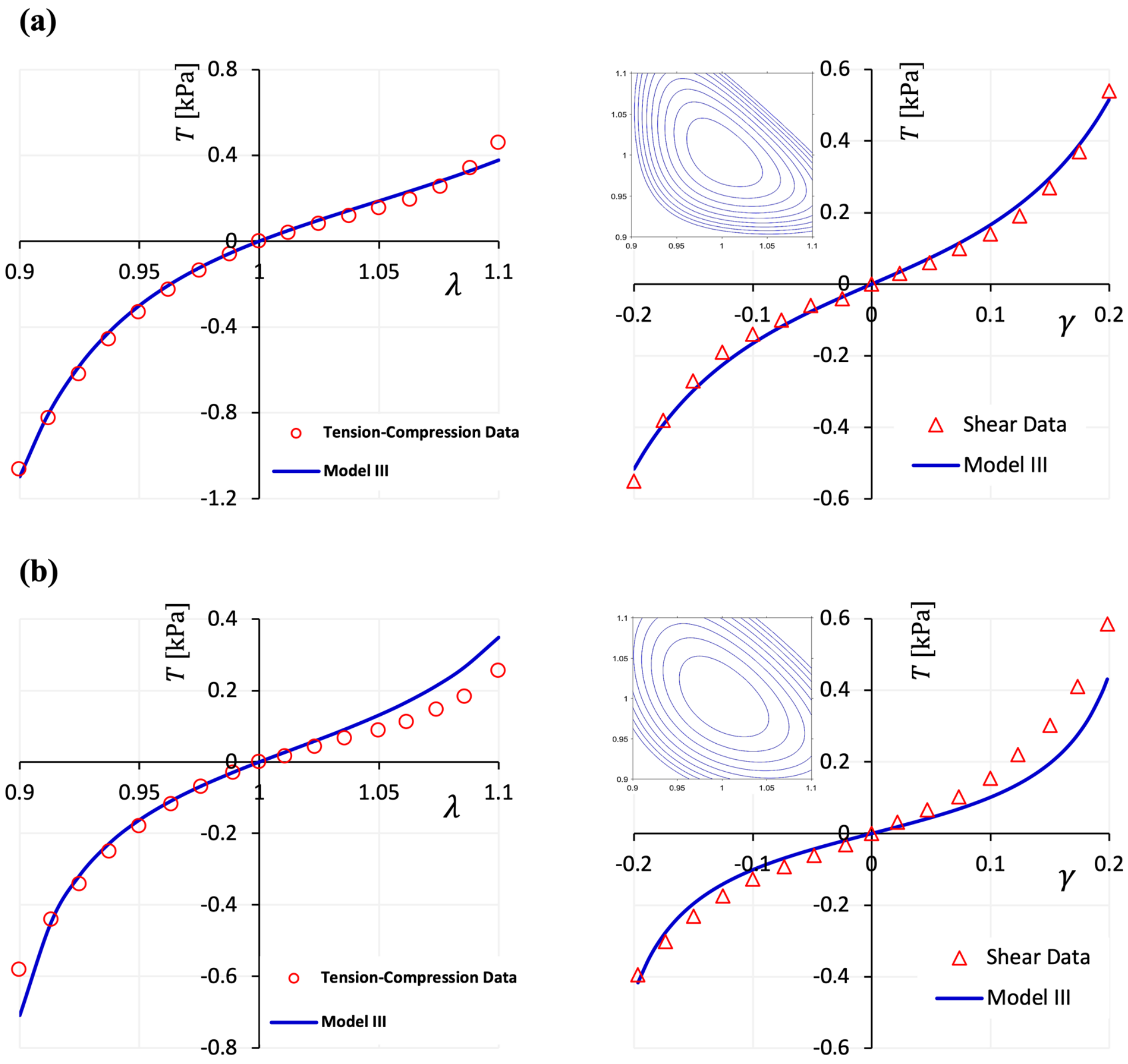

**Figure 8** – Fitting results of the *Model III* in equation (4.3) to the uniaxial compression-tension and simple shear datasets of human brain cortex due to: (a) Budday et al. [4]; and (b) Budday et al. [3]. The relationships used for the fittings are given by equation (4.4). Note the improved quality of fits even compared with the one-term Ogden model as presented in figures 2 and 4 for the same datasets, while maintaining convexity (insets).

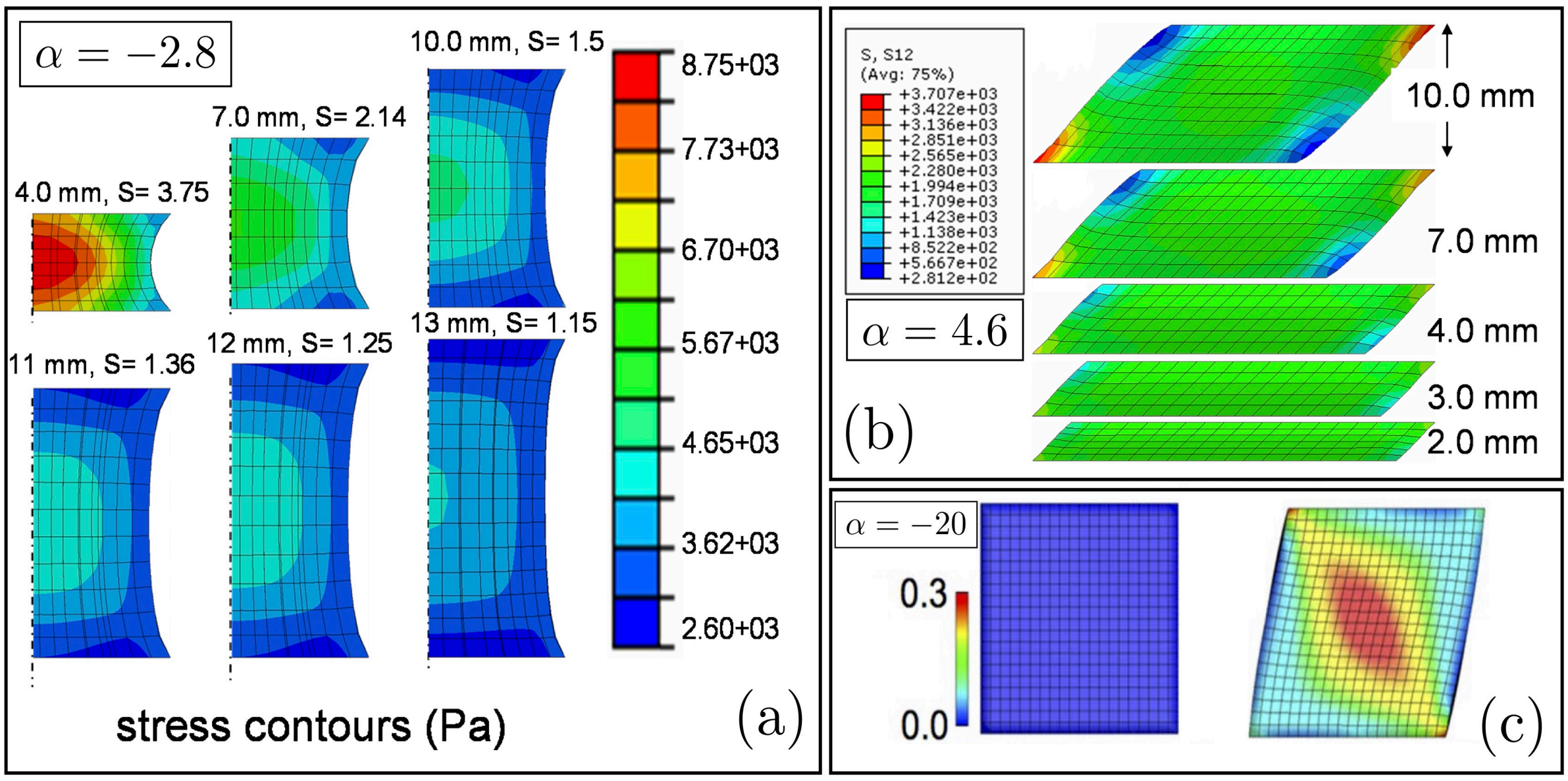

**Figure 9** – Finite Element simulations demonstrating the effect of inhomogeneity of stress distribution on stress predictions: (a) Uniaxial tension [44] for a cylinder of brain matter with diameter 15 mm, various heights, and Ogden parameter $\alpha = -2.8$. Stretch is 30% As the height/width ratio increases, the inhomogeneity effects become less marked; (b) Simple shear [6] for a rectangle of brain matter with width 19 mm, various heights, and Ogden parameter $\alpha = 4.6$. Amount of shear is $\gamma = 1.0$; and (c) Simple shear [1] for a square of brain matter with Ogden parameter $\alpha = -20$. Amount of shear is $\gamma = 0.2$.